\documentclass[conference,9pt]{IEEEtran}
\IEEEoverridecommandlockouts
\usepackage{cite}
\usepackage{amsmath,amssymb,amsfonts}
\usepackage{algorithmic}
\usepackage{graphicx}
\usepackage{textcomp}
\usepackage{xcolor}
\usepackage{ulem}
\usepackage{booktabs}
\def\BibTeX{{\rm B\kern-.05em{\sc i\kern-.025em b}\kern-.08em
    T\kern-.1667em\lower.7ex\hbox{E}\kern-.125emX}}
\begin{document}

\definecolor{note_color}{rgb}{1.0, 0.0, 0.0}
\newcommand{\TODO}[1]{\textcolor{note_color}{#1}}

\providecommand{\celine}[1]{{\protect\color{olive}{[celine: #1]}}}

\title{\fontsize{22}{12}\selectfont Spec2RTL-Agent: Automated Hardware Code Generation from Complex Specifications Using LLM Agent Systems}

\author{
\makebox[\textwidth][c]{
\begin{tabular}{c c c}
Zhongzhi Yu & Mingjie Liu & Michael Zimmer \\
\textit{Nvidia Research} & \textit{Nvidia Research} & \textit{Cadence} \\
Austin, TX & Austin, TX & San Jose, CA \\
zhongzhiy@nvidia.com & mingjiel@nvidia.com & zimmerm@cadence.com 
\\
\\
 Yingyan (Celine) Lin & Yong Liu & Haoxing Ren \\
 \textit{Georgia Institute of Technology} & \textit{Cadence} & \textit{Nvidia Research} \\
 Atlanta, GA & San Jose, CA & Austin, TX \\
 celine.lin@gatech.edu & yongl@cadence.com & haoxingr@nvidia.com
\end{tabular}
}
}

\maketitle

\begin{abstract}
    Despite recent progress in generating hardware register transfer level (RTL) code with large language models (LLMs), existing solutions still suffer from a substantial gap between practical application scenarios and the requirements of real-world RTL code development. Prior approaches either focus on overly simplified hardware descriptions or depend on extensive human guidance to process complex specifications, limiting their scalability and automation potential. In this paper, we address this gap by proposing an LLM agent system, termed Spec2RTL-Agent, designed to directly process complex specification documentation and generate corresponding RTL code implementations, advancing LLM-based RTL code generation toward more realistic application settings. To achieve this goal, Spec2RTL-Agent introduces a novel multi-agent collaboration framework that integrates three key enablers: (1) a reasoning and understanding module that translates specifications into structured, step-by-step implementation plans; (2) a progressive coding and prompt optimization module that iteratively refines the code across multiple representations (pseudocode, Python, and C++) to enhance correctness and synthesisability for RTL conversion; and (3) an adaptive reflection module that identifies and traces the source of errors during generation, ensuring a more robust code generation flow. Instead of directly generating RTL from natural language, our system strategically generates synthesizable C++ code, which is then optimized for high-level synthesis (HLS). This agent-driven refinement ensures greater correctness and compatibility compared to naive direct RTL generation approaches. We evaluate Spec2RTL-Agent on a benchmark of three specification documents, demonstrating its effectiveness in generating accurate RTL code with as much as 75\% fewer human interventions compared to existing approaches. These results underscore Spec2RTL-Agent’s role as the first fully automated multi-agent system for RTL generation from unstructured specification documents, reducing the reliance on human effort and expertise in hardware design.
\end{abstract}

\section{Introduction}
In recent years, Large Language Models (LLMs) have demonstrated remarkable performance across a diverse range of tasks, fundamentally transforming human activities and work pipelines~\cite{vicuna2023,gpt35,githubcopilot}. Given the labor-intensive nature of hardware code implementation, particularly at the Register Transfer Level (RTL), there is a compelling motivation to leverage advanced LLM technologies~\cite{fu2023gpt4aigchip,liu2023verilogeval,lu2023rtllm}. Deploying LLMs for RTL code generation offers a promising approach to enhance design automation productivity, reduce reliance on extensive human intervention, and accelerate the development cycle~\cite{blocklove2023chip,he2023chateda,lu2023rtllm,huang2024towards}.  

\textcolor{black}{However, existing LLM-based RTL generation approaches struggle to handle real-world hardware design tasks. Hardware specifications are often lengthy, contain unstructured multi-modal information (e.g., tables, figures, equations), and require deep contextual understanding beyond simple text-based instructions. Prior works, such as VerilogEval~\cite{liu2023verilogeval} and GPT4AIGChip~\cite{fu2023gpt4aigchip}, focus on either generating isolated RTL components or simplifying the specification input, thereby limiting their applicability to complex, multi-stage hardware designs. For instance, generating an RTL implementation for an encryption module requires not only extracting constraints and defining submodules but also iteratively refining the design based on verification—tasks that existing single-pass LLM methods fail to automate.  }

\begin{figure}
    \centering
    \includegraphics[width=\linewidth]{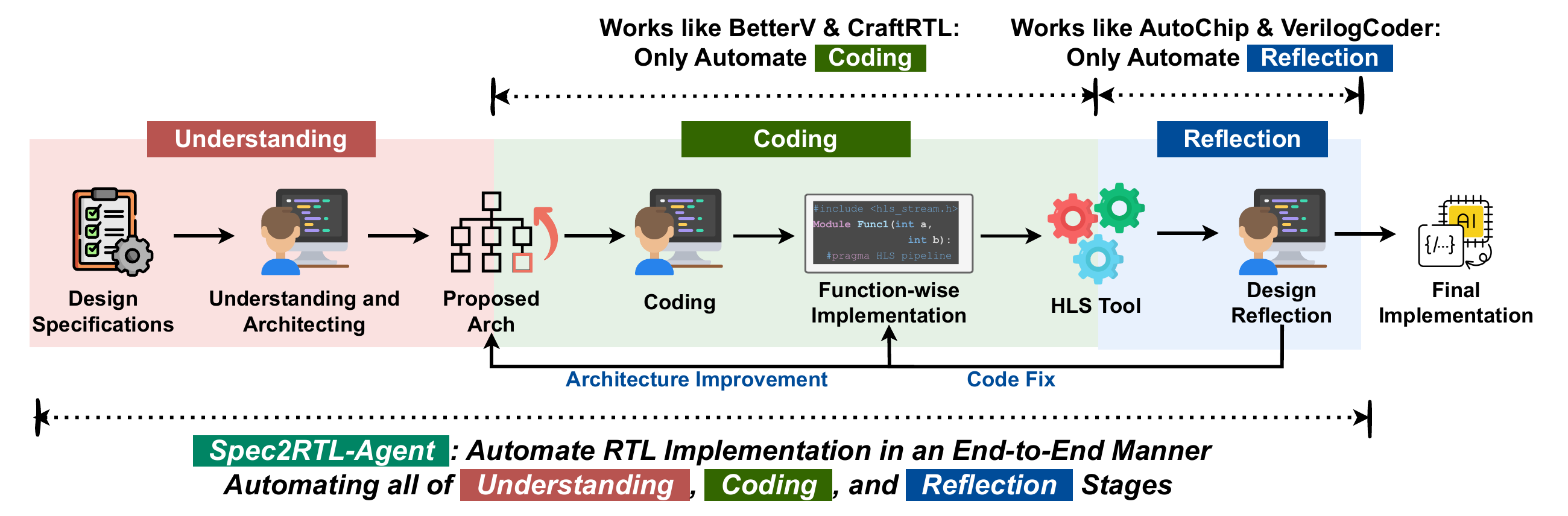}
    \vspace{-2em}
    \caption{Overview of the hardware implementation flow and the comparison between the capability of existing LLM-based approaches in automating hardware code generation and the achieved automated hardware code generation with our proposed Spec2RTL-Agent.
    }
    \label{fig:motivation}
    \vspace{-1em}
\end{figure}

\textcolor{black}{As illustrated in Fig.~\ref{fig:motivation}, current LLM-based RTL generation approaches can be categorized into two groups: (1) methods that target simplified scenarios, such as generating standalone RTL functions from structured natural language descriptions~\cite{liu2023verilogeval,lu2023rtllm,liu2024craftrtl,ho2024verilogcoder}, or automating only specific sub-parts of the hardware design pipeline~\cite{autodnnchip,fu2023gpt4aigchip}, and (2) approaches that integrate LLMs into human-in-the-loop workflows, where engineers remain responsible for specification interpretation, task decomposition, and iterative debugging~\cite{wang2024chatcpu,blocklove2023chip}. While the latter category benefits from human expertise, it still demands extensive manual intervention, preventing full automation.}

To address these limitations, we propose \textbf{Spec2RTL-Agent}, an LLM-based multi-agent system designed to automate RTL code generation directly from unstructured specification documents. Unlike previous approaches that either generate RTL directly or rely on extensive human preprocessing, Spec2RTL-Agent follows a structured, iterative process that mirrors the conventional human-driven hardware design workflow. Our framework systematically processes specification documents, refines code generation through multiple abstraction levels, and iteratively verifies outputs to improve correctness.

\textcolor{black}{Inspired by the way human engineers approach hardware design, Spec2RTL-Agent operates in three critical stages: (1) \textbf{Understanding}, analyzing the specification to extract constraints and structure implementation plans, ensuring clarity and correctness from the outset, (2) \textbf{Coding}, progressively refining the implementation by first generating structured high-level code before transforming it into synthesizable C++ for High-Level Synthesis (HLS), and (3) \textbf{Reflection}, iteratively verifying, debugging, and refining intermediate outputs through an adaptive refinement mechanism to enhance robustness. This structured workflow aligns with traditional hardware development methodologies~\cite{nao_hardware_stages,encata_hardware_stages,dynedge_hardware_development}, allowing LLM agents to not only generate code but also engage in high-level reasoning, optimization, and error correction, reducing reliance on human intervention.}

The contributions of this paper are summarized as follows:
\begin{itemize}
    \item We introduce the first LLM-based system, dubbed the Spec2RTL-Agent, that emulates the human hardware implementation process to enable end-to-end RTL code generation directly from real-world complex specification documents.  This development marks a significant advancement in LLM-based RTL code generation, moving towards more practical and realistic applications by enhancing the productivity of hardware implementations and facilitating a more agile development process.

    \item Spec2RTL-Agent integrates three key enablers, each replicating a critical stage of the human-driven RTL implementation process, including (1) an \textbf{iterative understanding and reasoning module} to enhance the comprehension of documentation and facilitate strategic implementation planning, (2) a \textbf{progressive coding and prompt optimization module} that improves the efficiency and effectiveness of the coding process, and (3) an \textbf{adaptive reflection module} that enables adaptively debugging throughout the whole implementation flow, ensuring a robust and reliable code generation process.

    \item Experiments conducted on a benchmark containing three representative specification documents published by the National Institute of Standards and Technology (NIST) and their corresponding implementations validate the effectiveness of our proposed Spec2RTL-Agent in directly generating RTL implementations from specification documentation with limited human intervention. Notably, Spec2RTL-Agent achieves as much as 75\% fewer human interventions compared to existing methods. These results underscore Spec2RTL-Agent as a practical solution for real-world RTL code generation applications, significantly reducing the human labor typically required in the hardware implementation process.
\end{itemize}

\section{Background}

\subsection{LLM-Based RTL Code Generation}
\textcolor{black}{Recent work has applied LLMs to RTL code generation, enabling synthesis from textual descriptions~\cite{deng2025scalertl,liu2023verilogeval,lu2023rtllm,zhang2024mg,he2023chateda,liu2024craftrtl,liu2023chipnemo}. Methods such as ChipAlign~\cite{deng2024chipalign}, MG-Verilog~\cite{zhang2024mg}, and CraftRTL~\cite{liu2024craftrtl} generate RTL directly but struggle with unstructured inputs, requiring human oversight for refinement and validation. RTLCoder~\cite{liu2024rtlcoder} improves generation quality, achieving state-of-the-art performance among non-commercial tools. Other studies integrate LLMs into human-in-the-loop workflows~\cite{blocklove2023chip,wang2024chatcpu,fu2023gpt4aigchip}, where engineers iteratively guide the model, enhancing correctness but limiting scalability. In contrast, Spec2RTL-Agent reduces human intervention through structured reasoning and iterative refinement within a multi-agent framework.}

\textcolor{black}{Another line of work targets synthesizable C++ for HLS, offering a higher level of abstraction~\cite{xiong2024hlspilot,swaroopa2024evaluating}. However, these methods assume well-structured inputs and lack iterative correction, making them less suitable for raw specifications. Spec2RTL-Agent improves automation robustness by refining intermediate representations prior to synthesis.}

\subsection{LLM Agents}
Recent advances in LLMs have enabled multi-agent systems for complex tasks such as code generation, mathematical proof, and story writing~\cite{chen2023autoagents,yuan2024evoagent,wuautogen,talebirad2023multi,wang2307unleashing}. Multi-agent collaboration has been shown to outperform single-agent reasoning~\cite{chen2023autoagents,wang2307unleashing}. AgentCoder~\cite{huang2023agentcoder} iteratively refines code through agent interactions, and AutoSafeCoder~\cite{nunez2024autosafecoder} enhances security with static analysis and fuzz testing. In the RTL domain, VerilogCoder~\cite{ho2024verilogcoder} introduces graph-based planning and syntax-tree tracing, while ChatCPU~\cite{wang2024chatcpu} uses human-LLM collaboration to accelerate complex designs. These methods still rely heavily on human input, motivating our development of a fully autonomous LLM-agent system for RTL code generation.

\subsection{Automated Hardware Design Methods}
\textcolor{black}{Beyond LLMs, RTL design has been explored via heuristic and search-based techniques, including genetic algorithms and reinforcement learning~\cite{sohrabizadeh2022autodse,alur2013syntax,chen2020hardware}. MCTS has been combined with LLMs to address synthesis failures and improve power, performance, and area (PPA) metrics~\cite{delorenzo2024make}. However, such methods typically depend on structured constraints and fixed inputs. Spec2RTL-Agent instead couples language-driven reasoning with iterative refinement, enabling robust automation from unstructured specifications.}

\section{Preliminary}
\label{sec:preliminary}
\begin{figure}
    \centering
    \includegraphics[width=0.8\linewidth]{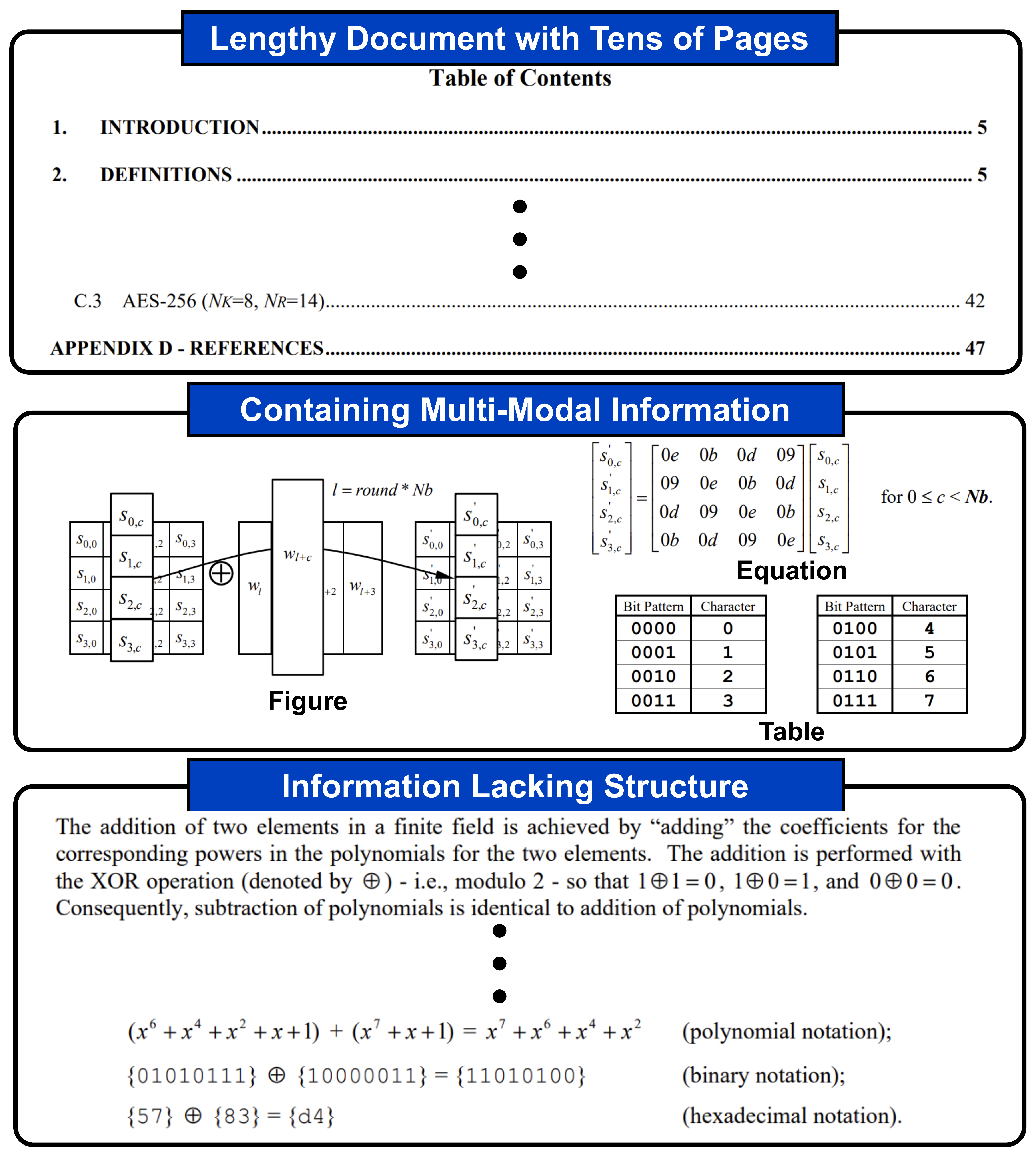}
    \vspace{-1em}
    \caption{Visualization of a representative specification document used as input for our target problem. The expected specification document consists of multi-modal information (e.g., tables, figures, and equations) in a lengthy and unstructured manner. }
    \label{fig:document_vis}
    \vspace{-1em}
\end{figure}

\subsection{Problem Definition}
\label{sec:problem}
\textcolor{black}{The goal of this work is to develop an end-to-end framework that generates functionally correct RTL implementations directly from hardware specification documents while minimizing human intervention. Unlike structured benchmarks such as VerilogEval~\cite{liu2023verilogeval}, real-world hardware specifications contain unstructured, multi-modal information, making them difficult for existing LLM-based approaches to process effectively. These methods either rely on structured inputs or operate in a single-pass manner, leading to errors that require extensive human debugging~\cite{wang2024chatcpu,ho2024verilogcoder}.}

We define the key aspects of this problem as follows:
\begin{itemize}
    \item \textbf{Goal}: The system should automate RTL generation from raw specification documents, significantly reducing human effort in hardware design while ensuring correctness and consistency.

    \textbf{Input:} A hardware specification document in its original form, without manual simplifications. These documents typically contain extensive, unstructured content, including figures, tables, and equations (Fig.~\ref{fig:document_vis}), making them fundamentally different from structured prompts used in prior LLM-based RTL generation benchmarks.
    
    \textbf{Output:} A functionally correct RTL implementation that satisfies the constraints and functional requirements specified in the input document.
    
    \textbf{Expected Level of Automation:} The system should autonomously perform the majority of RTL development tasks, requiring minimal human intervention, with each intervention limited to high-level guidance or addressing cases beyond the model’s reasoning capabilities. Additionally, it should summarize key issues encountered during the process, reducing the burden on human developers and minimizing the need for extensive manual corrections across multiple iterations.

\end{itemize}

\subsection{Identified Challenges}
\label{sec:challenge}
Given the capability of current LLMs, we identify the following challenges that must be addressed to tackle the problem defined above:
\begin{itemize}
    \item \textbf{\uline{Challenge 1}: Understanding and Reasoning}. A fundamental challenge is the comprehensive understanding of complex, unstructured, and multi-modal information presented in specification documents and translating this into structured, actionable implementation plans. Current LLM-based approaches typically perform a single LLM inference pass, which is insufficient for handling the intricacies of such data, resulting in a heavy reliance on human intervention for guidance during the implementation process~\cite{chang2023chipgpt,blocklove2023chip,wang2024chatcpu}.
    
    \item \textbf{\uline{Challenge 2}: Multi-Module Code Generation}: The complexity of generating complete RTL implementations in a single attempt exceeds the current capabilities of LLMs, particularly due to their limitations in producing long output sequences~\cite{xiao2023efficient,xiao2024duoattention}. Therefore, it is essential to develop a method that enables controlled, stepwise generation of multi-module implementations, while ensuring logical consistency and seamless integration across these modules.
    
    \item \textbf{\uline{Challenge 3}: Error Detection and Correction}: The multi-module and multi-step generation approach significantly complicates error detection, a critical aspect of the RTL implementation process. This is because errors may not only arise from the current segment of the code but also from previously implemented functions or even the initial implementation plan. Existing methods typically focus on error debugging within simplified, single-module scenarios~\cite{xu2024meic,ho2024verilogcoder,huang2024towards}, proving inadequate for managing the complexities inherent in comprehensive RTL designs.
\end{itemize}
\section{The Proposed Spec2RTL-Agent System}

\begin{figure*}
    \centering
    \includegraphics[width=\textwidth]{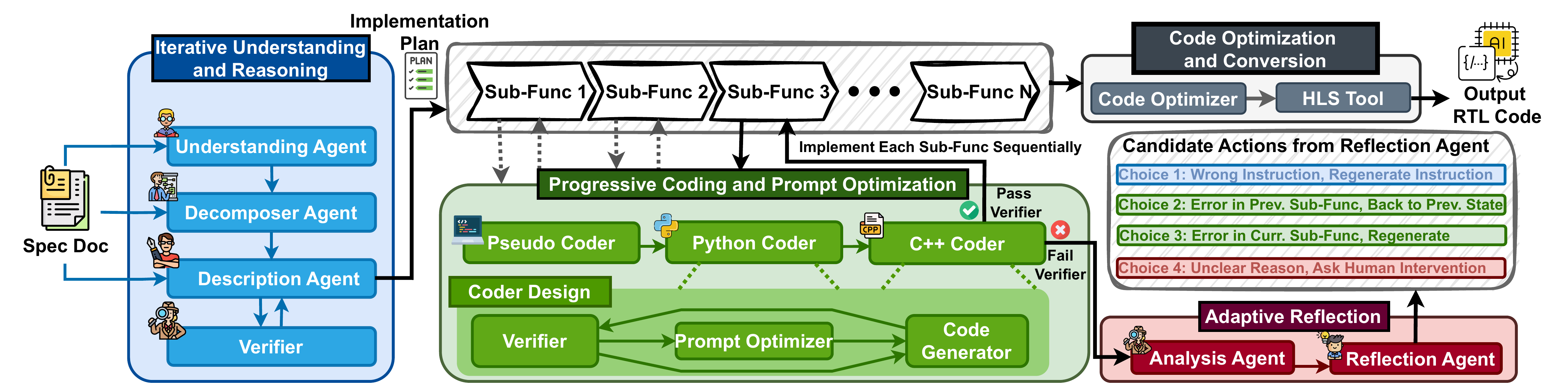}
    \vspace{-2em}
    \caption{Overview of our proposed Spec2RTL-Agent system. }
    \label{fig:overview}
    \vspace{-1em}
\end{figure*}
\subsection{Overview}
To address the challenges outlined in Sec.~\ref{sec:challenge}, we propose \textbf{Spec2RTL-Agent}, a novel multi-agent framework designed to automate RTL generation directly from complex, unstructured specification documents. Unlike prior methods that either generate RTL in a single step~\cite{liu2023verilogeval,ho2024verilogcoder} or assume well-structured high-level descriptions for HLS~\cite{xiong2024hlspilot}, Spec2RTL-Agent introduces an \textit{iterative} processing pipeline that systematically refines RTL implementations through multi-stage reasoning, code generation, and error correction.

As illustrated in Fig.~\ref{fig:overview}, Spec2RTL-Agent operates through three key enablers:
\begin{enumerate}
    \item \textbf{Iterative Understanding and Reasoning Module} (Sec.~\ref{sec:understand}): Initially, the specification document undergoes analysis by the iterative understanding and reasoning module, which features a propose-and-verify-based collaboration document understanding and reasoning pipeline. This module systematically decomposes the target function into multiple sub-functions, each formatted in an LLM-friendly manner to effectively address \textbf{\uline{Challenge 1}: Understanding and Reasoning}.
    
    \item \textbf{Progressive Coding and Prompt Optimization Module} (Sec.~\ref{sec:coding}): Following the plan, the system sequentially implements each sub-function using the progressive coding and prompt optimization module. This module utilizes cross-level code referencing and prompt revision techniques to enable a structured and efficient multi-module implementation, thereby alleviating \textbf{\uline{Challenge 2}: Multi-Module Code Generation}.
    \item \textbf{Adaptive Reflection Module} (Sec.~\ref{sec:reflect}): As module implementation progresses, the limited availability of test cases for each sub-function complicates the verification of correctness. The adaptive reflection module dynamically identifies and addresses sources of errors across all generated sub-functions and even within the instructions, aiming to ensure a robust and error-free generation process as outlined in \textbf{\uline{Challenge 3: Error Detection and Correction}}.
    \item \textbf{Code Optimization and Conversion} (Sec.~\ref{sec:conversion}): Upon completing all sub-functions, a code optimization agent reformats the implementation for compatibility with the HLS tool. The system then utilizes the HLS tool to convert the processed implementation into the target RTL code.
\end{enumerate}
We will introduce each key module in the following sections.

\subsection{Iterative Understanding and Reasoning Module}
\label{sec:understand}
The iterative understanding and reasoning module is designed to transform the original, lengthy, unstructured, and LLM-unfriendly specification document into a concise, structured, and LLM-friendly implementation plan. This transformation is facilitated by a multi-agent collaboration pipeline, which iteratively verifies and refines the content to gradually convert and organize the necessary information in a structured format. The process within this module consists of three stages:
\begin{itemize}
    \item \textbf{Summarization}: Firstly, we aim to alleviate the burden on subsequent agents in directly processing the entirety of the lengthy specification document by condensing the information contained in each section. A Summarization Agent is employed to generate concise summaries for each section of the document, thereby simplifying the initial comprehension process and reducing the complexity for further processing.
    
    \item \textbf{Decomposition}: Building on the summarized information, we initiate the reasoning process by breaking down the complex target function into manageable sub-functions. A Decomposer Agent receives both the summarized data and the original document, tasked with organizing the target implementation into a sequence of implementable sub-functions.
    
    \item \textbf{Information Augmentation}: The final stage aims to supplement each decomposed sub-function with the necessary details to facilitate direct implementation by subsequent modules, without the need to refer back to the original document. This task is accomplished through the cooperative efforts of a Description Agent and a Verifier.The Description Agent uses the original document, the summaries provided, and specific sub-function requirements as inputs to collate and format the necessary information into a structured dictionary. This dictionary includes key elements such as inputs, outputs, functionality, and pertinent references from the original document. The Verifier then reviews this dictionary, providing feedback to enhance the quality and accuracy of the information. This collaborative approach ensures that each sub-function is equipped with complete and precise implementation details.
\end{itemize}
Upon completion of these stages, the iterative understanding and reasoning module forwards the refined decomposition plan and the associated information for each sub-module to the subsequent Progressive Coding and Prompt Optimization Module to start code implementation. 

\subsection{Progressive Coding and Prompt Optimization Module}
\label{sec:coding}
The progressive coding and prompt optimization module is responsible for implementing the code based on the decomposition plan and associated information generated from the iterative understanding and reasoning module. This module sequentially constructs each sub-function and integrates them into a comprehensive implementation. Within this module, we have developed two innovative techniques to enhance accuracy and efficiency: 
\begin{itemize}
    \item \textbf{Progressive Coding}: Inspired by the varying proficiency levels of LLMs in handling different programming languages, from high-level to low-level code, we leverage higher-level code as a reference to guide the generation of corresponding lower-level implementations. The process involves sequentially generating pseudocode, Python, and C++ code. For each sub-function, a Coder first drafts the code and a Verifier then assesses its accuracy, suggesting necessary revisions or a complete regeneration when needed. The Verifier either derives test cases from the specification document or utilizes those created from previously implemented higher-level code. In scenarios where external test cases are unavailable, the Verifier performs a self-validation by directly comparing the implementation against the original specifications. 
    \item \textbf{Prompt Optimization}: The iterative interaction between the Coder and Verifier, involving repeated code generation and revisions for each sub-function, may lead to inefficiencies. To streamline this process and enhance cost-effectiveness, we introduce a Prompt Optimizer Agent. This agent analyzes the implementation log for the current sub-function, extracts learnings, and refines the prompts used by the Coder to enhance the accuracy and efficiency of subsequent coding attempts.
\end{itemize}

\subsection{Adapative Reflection Module}
\label{sec:reflect}
Despite the effectiveness of the progressive coding and prompt optimization module, the absence of sub-function-specific test cases can hinder the accurate identification of implementation errors. This often results in scenarios where a newly implemented sub-function fails to pass test cases due to errors in previously implemented sub-functions. To address this challenge and enhance the robustness and error-tolerance of the implementation pipeline, we introduce the adaptive reflection module.

The adaptive reflection module operates in two primary stages:
\begin{itemize}
    \item \textbf{Error Source Analysis}: An Analysis Agent first reviews the entire generation trajectory to summarize the completed work and propose potential areas where errors may reside.
    \item \textbf{Error Resolution Direction}: Following this analysis, a Reflection Agent evaluates the summarized information and potential error sources. It then determines the most appropriate course of action from four possible strategies:
    \begin{itemize}
        \item If the error originates from \textit{incorrect instructions}, the agent redirects the issue to the iterative understanding and reasoning module to revise the instructions for the affected sub-function.
        \item If the error stems from \textit{previous sub-functions}, the Reflection Agent returns to the identified sub-functions to make necessary revisions, using the current failing test cases as guidance.
        \item If the error is confined to the \textit{current sub-function} and unrelated to previous outputs, the Reflection Agent restarts the generation of this sub-function within the progressive coding and prompt optimization module, providing targeted feedback to the Coder to avoid repeating the error.
        \item If the error source \textit{remains unclear}, the agent escalates the issue for human intervention, summarizing the situation and requesting guidance on corrective measures.
    \end{itemize}
\end{itemize}

\subsection{Code Optimization and Conversion Module}
\label{sec:conversion}
Upon the completion of all sub-functions as per the implementation plan, the next critical step involves converting the implemented C++ code into RTL using a leading commercial HLS tool, Stratus High-Level Synthesis (Stratus HLS)~\cite{stratus_hls}. To ensure compatibility with the HLS tool, which has specific coding format requirements, we introduce a Code Optimizer Agent. This agent adapts the C++ code to meet the HLS tool's constraints based on guidelines extracted from the tutorial of this HLS tool. Key adaptations include conforming to data format restrictions and optimizing for static memory usage. The Code Optimizer Agent systematically applies these rules to prepare the code for efficient and error-free high-level synthesis.
\section{Experiments}

\subsection{Evaluation Setting}
\uline{Benchmark}: To evaluate the effectiveness of our Spec2RTL-Agent framework, we have chosen a subset of three standard specification documents from the Federal Information Processing Standards (FIPS) developed by NIST. These documents include the Advanced Encryption Standard (AES)~\cite{aes}, Digital Signature Standard (DSS)~\cite{dss}, and Keyed-Hash Message Authentication Code (HMAC)~\cite{HMAC}. This selection aims to provide a robust evaluation across a set of representative RTL generation scenarios.

\begin{table}[]
\vspace{-1em}

    \centering
    \caption{Benchmarking Spec2RTL-Agent with baseline solutions on different FIPS documents.}
    \vspace{-1em}
    \resizebox{\linewidth}{!}{
    \begin{tabular}{cccccccc}\toprule
Method &Human &Single-Shot &W/o Understand &Naive Coding &W/o Reflection &Spec2RTL-Agent \\\cmidrule{1-7}
Correct & 3/3 & 0/3 & 2/3 & 3/3 & 3/3 & 3/3 \\\cmidrule{1-7}
\# Intervention &$\sim$20 &/ &18.67 &9.00 &6.33 &4.33 \\\cmidrule{1-7}
\# Coding &$\sim$20 &/ &15.53 &17.46 &13.20 &9.11 \\
\bottomrule
\end{tabular}
}
\vspace{-1.5em}
    \label{tab:main_exp}
\end{table}

\uline{Evaluation Metrics}: To comprehensively assess the performance of our RTL generation system, we consider the following metrics: (1) \textbf{Correct}: Functional correctness of the generated RTL implementation, (2) \textbf{\# Intervention}: Number of human interventions required to achieve correct functionality, and (3) \textbf{\# Coding}: Average number of code generation and revision attempts needed per sub-function.

\uline{Baselines}: Given that our proposed Spec2RTL-Agent is a pioneering system for automating end-to-end RTL generation from specification documents, direct baselines are scarce. Therefore, we have established the following baselines to benchmark its performance:
(1) \textbf{Single-Shot}: The entire specification document is input into an LLM, which is tasked with generating the target RTL in a single attempt.
(2) \textbf{Human}: Mirroring approaches like that in~\cite{blocklove2023chip}, a human handles the understanding and reflection phases while delegating code implementation to an LLM based on provided instructions.
(3) \textbf{W/o Understand}: This variation of our pipeline omits the iterative understanding and reasoning module, directly feeding the entire document into the coding module to complete the implementation.
(4) \textbf{Naive Coding}: This configuration removes the progressive coding and prompt optimization module, replacing it with a naive C++ coding module. This module directly generates the C++ code for each target sub-function using the information provided, without iterative enhancements or optimizations.
(5) \textbf{W/o Reflection}: This configuration bypasses the adaptive reflection module. Errors must be addressed by the LLM within the current sub-function without the capability to revisit previously implemented sub-functions or instructions.

\uline{Implementation Details}: In our study, we use GPT-4o~\cite{gpt4} as the core architecture for all agents, with agent interactions and tool usage managed via the AutoGen framework~\cite{wuautogen}. Due to LLMs' limitations in processing PDFs, we extract text using PyPDF and capture screenshots for figures and tables. All extracted data is compiled and fed into the LLM. During code generation, the full target code often exceeds LLMs’ generation limits, making it impractical to regenerate the entire file for every sub-function update. To address this, agents generate only the modified sub-function, mark it with 20 asterisks (*), and label its start with the sub-function name. A rule-based approach then locates and replaces the corresponding sub-function in the code file.  We use Stratus HLS~\cite{stratus_hls} as our HLS tool.

\subsection{Benchmarking Spec2RTL-Agent on Specification Documents} 
We evaluated Spec2RTL-Agent against baseline approaches using selected benchmark specification documents. As shown in Table~\ref{tab:main_exp}, Spec2RTL-Agent consistently outperformed baselines across various metrics. The \textbf{Single-Shot} approach, which directly prompts LLMs to generate implementations, failed in all three test cases, underscoring the complexity of the task. In contrast, Spec2RTL-Agent transformed an otherwise ineffective process into an efficient pipeline for generating complex RTL implementations.

Regarding human effort reduction, Spec2RTL-Agent significantly lowered the need for intervention compared to the \textbf{Human} baseline, where humans handle understanding, reasoning, and reflection while LLMs only implement code~\cite{blocklove2023chip}. Specifically, Spec2RTL-Agent reduced human interventions by approximately 75\%, easing the burden of RTL implementation and enhancing development efficiency.

A detailed performance breakdown across specification documents is shown in Table~\ref{tab:breakdown}. Despite document diversity, Spec2RTL-Agent effectively implemented RTL with around five human interventions and fewer than ten code revision attempts per sub-function. To further validate its RTL generation quality, we compared the achievable latency, throughput, and area of our generated AES engine with modified open-source AES solutions, ensuring HLS compatibility. Using Stratus HLS~\cite{stratus_hls} across multiple configurations, our solution achieved comparable performance metrics.

\begin{table}[]
\vspace{-1em}
    \centering
    \caption{Performance of Spec2RTL-Agent across different documents.}
    \vspace{-1em}
    \resizebox{0.6\linewidth}{!}{\begin{tabular}{cccc}\toprule
Function &AES &DSS &HMCA \\\cmidrule{1-4}
\# Intervention &4 &6 &3 \\\cmidrule{1-4}
\# Coding &8.49 &9.31 &9.52 \\
\bottomrule
\end{tabular}
}
    \label{tab:breakdown}
    \vspace{-1.5em}
\end{table}

\subsection{Ablation Study on the Effectiveness of Each Module}
To assess the contributions of individual modules in Spec2RTL-Agent, we benchmarked against three baselines, each omitting a system module. As shown in Table~\ref{tab:main_exp}, removing the iterative understanding and reasoning module (\textbf{W/o Understanding}) led to a more than fourfold increase in human interventions, underscoring its importance in planning. Compared to \textbf{W/o Reflection} and \textbf{Naive Coding}, our full Spec2RTL-Agent reduced human interventions by 51.9\% and 31.6\%, and coding iterations by 47.8\% and 31.0\%, respectively. These results confirm the effectiveness of each module in improving code accuracy.

\begin{table}[b]
\vspace{-2em}
    \centering
    \caption{Ablate on robustness to generation noise.}
    \vspace{-1em}
    \resizebox{0.9\linewidth}{!}{\begin{tabular}{cccccc}\toprule
Noise Loc. & None & Understanding & Pseudocode & Python & C++  \\\cmidrule{1-6}
Correct & 3/3 & 3/3 & 3/3 & 3/3 & 3/3  \\\cmidrule{1-6}
\# Intervention &4.33 & 6.00 & 4.67 & 4.67 & 5.33 \\\cmidrule{1-6}
\# Coding &9.11 &11.42 &9.31 &  9.42 & 9.88\\
\bottomrule
\end{tabular}
}
    \label{tab:robustness}
    % \vspace{-1em}
\end{table}

\begin{figure}[t]
    \centering
    \includegraphics[width=\linewidth]{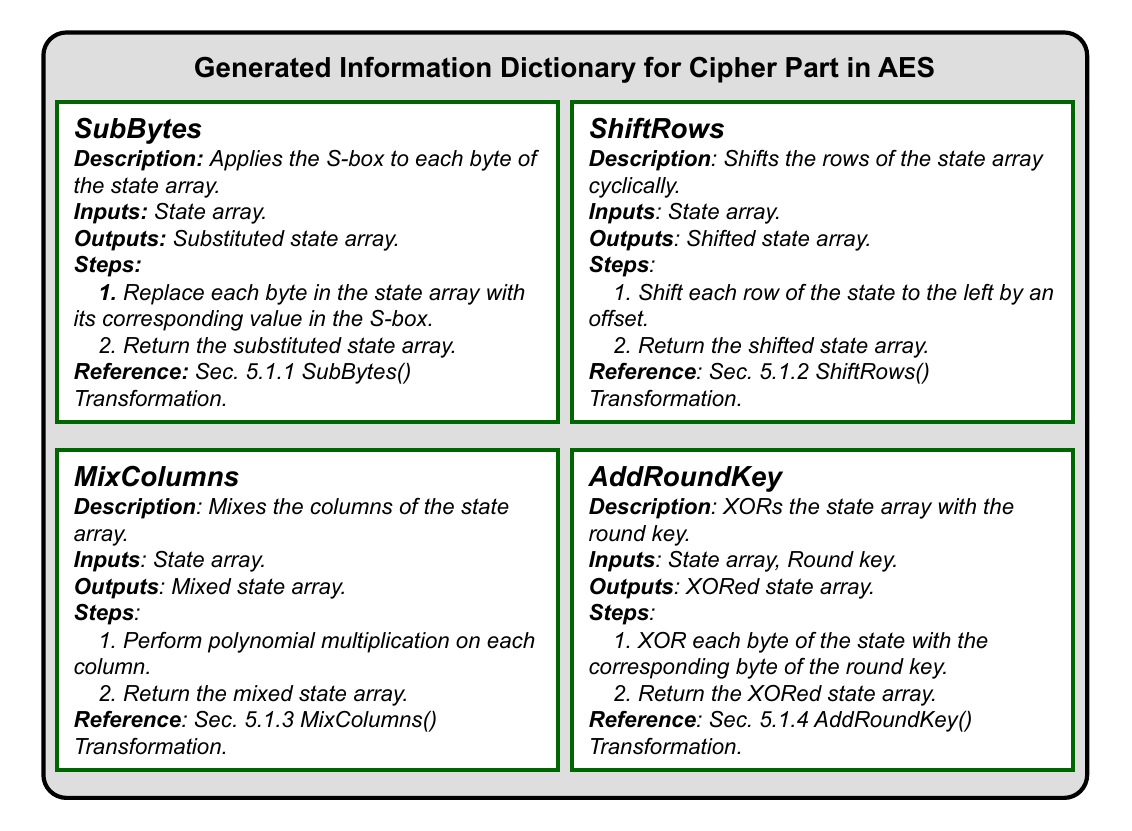}
      \vspace{-2em}
    \caption{Visualization of the generated information dictionary cipher function in the AES implementation.}
    \label{fig:information_dict}
         \vspace{-2em}

\end{figure}

\subsection{Ablation Study on the Robustness to Generation Noise}
\textcolor{black}{During the extended generation process, previously generated content may contain errors that propagate through subsequent stages. To assess Spec2RTL-Agent’s robustness against generation noise, we systematically introduce controlled semantic errors at different implementation levels. After each stage, we prompt the LLM to revise its output while embedding subtle but critical logical flaws: \textit{Revise the generated implementation while introducing a subtle but impactful mistake. The mistake should not be a syntax error but rather a semantic flaw that could lead to functional misbehavior. Ensure that the revised implementation remains coherent and well-structured.} We evaluate the impact of these errors across the implementation plan and three levels of code refinement. As shown in Table~\ref{tab:robustness}, Spec2RTL-Agent demonstrates strong resilience, consistently generating correct implementations while maintaining stable efficiency in \#Intervention and \#Coding metrics. These results highlight the system’s ability to recover from local inconsistencies and maintain high-quality output despite intentional perturbations.}

\subsection{Visualization on the Generation Process}
To elucidate the generation process of our proposed Spec2RTL-Agent, we present the following visualizations, offering insights into its operational workflow.

\textbf{Information Dictionary.} We visualize the generated information dictionary for sub-functions within the AES cipher function, as shown in Fig.~\ref{fig:information_dict}. Through the iterative understanding and reasoning module in Spec2RTL-Agent, unstructured and complex information from the specification document is transformed into structured representations. These representations explicitly define key attributes for each sub-function, such as inputs, outputs, and functionality, while referencing the original text. The effectiveness of this module is demonstrated in Table~\ref{tab:main_exp}, highlighting its role in organizing and clarifying essential details to enhance the generation process.

\begin{figure}[b]
\vspace{-2em}
    \centering
    \includegraphics[width=0.75\linewidth]{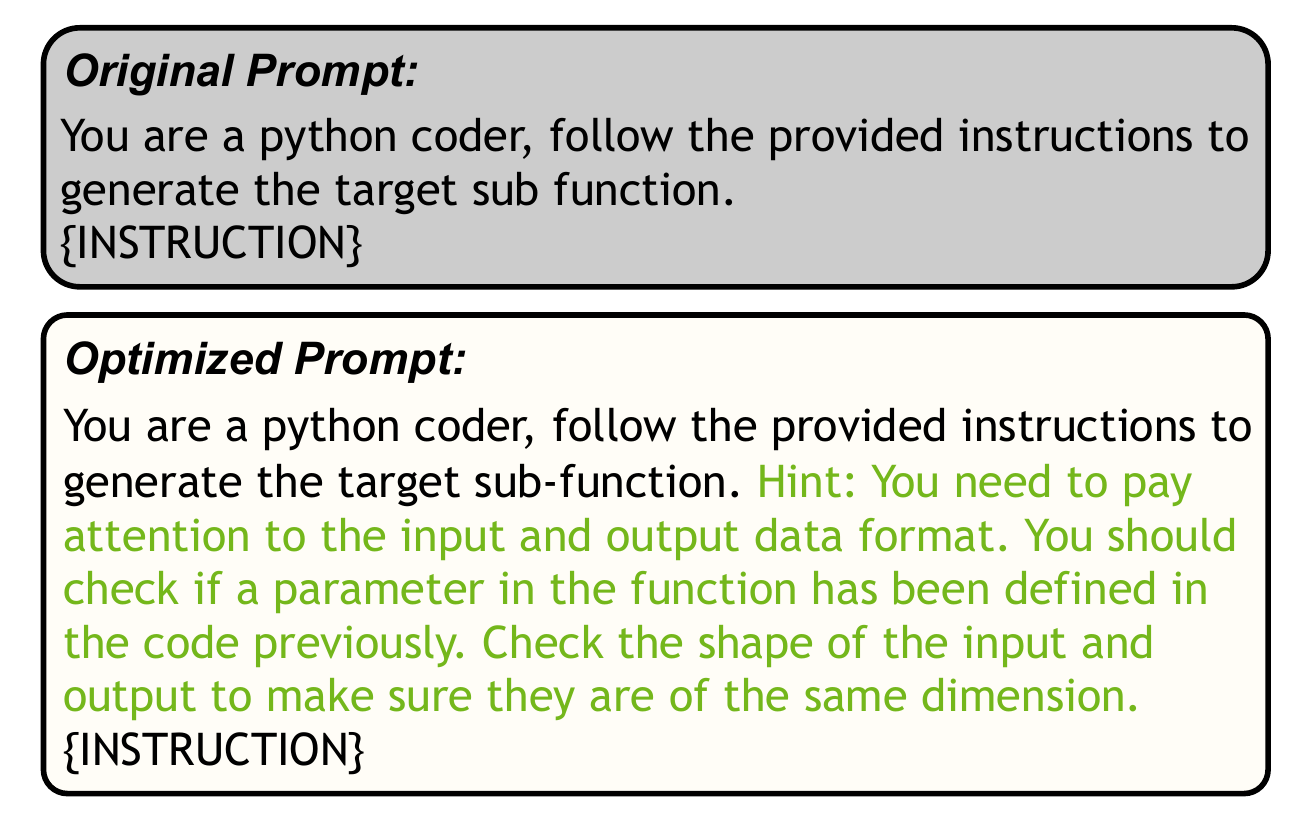}
     \vspace{-1.3em}
    \caption{Visualizing the prompt used to generate Python code for the \textit{AddRoundKey()} function before and after prompt optimization.}
    % \vspace{-1em}
    \label{fig:prompt_opt}
\end{figure}

\textbf{Prompt Optimization.} To illustrate the impact of prompt optimization, we depict its process and outcomes in Fig.~\ref{fig:prompt_opt}. Initially, a concise base prompt is supplemented with a sub-function-specific prompt (upper half of Fig.~\ref{fig:prompt_opt}) to generate sub-function implementations. However, repeated failures to pass test cases prompt the optimizer agent to iteratively refine its strategy by analyzing errors and adjusting the prompt accordingly. As shown in the lower half of Fig.~\ref{fig:prompt_opt}, the \textit{AddRoundKey()} function often exhibits interface mismatches with the higher-level cipher function, particularly in input-output dimensions. Recognizing these patterns, the optimizer integrates input-output compatibility details into the revised prompt, improving alignment with test case requirements. This adaptive approach demonstrates the system’s ability to enhance generation quality through iterative self-supervised adjustments.

\begin{figure}
    \centering
    \includegraphics[width=\linewidth]{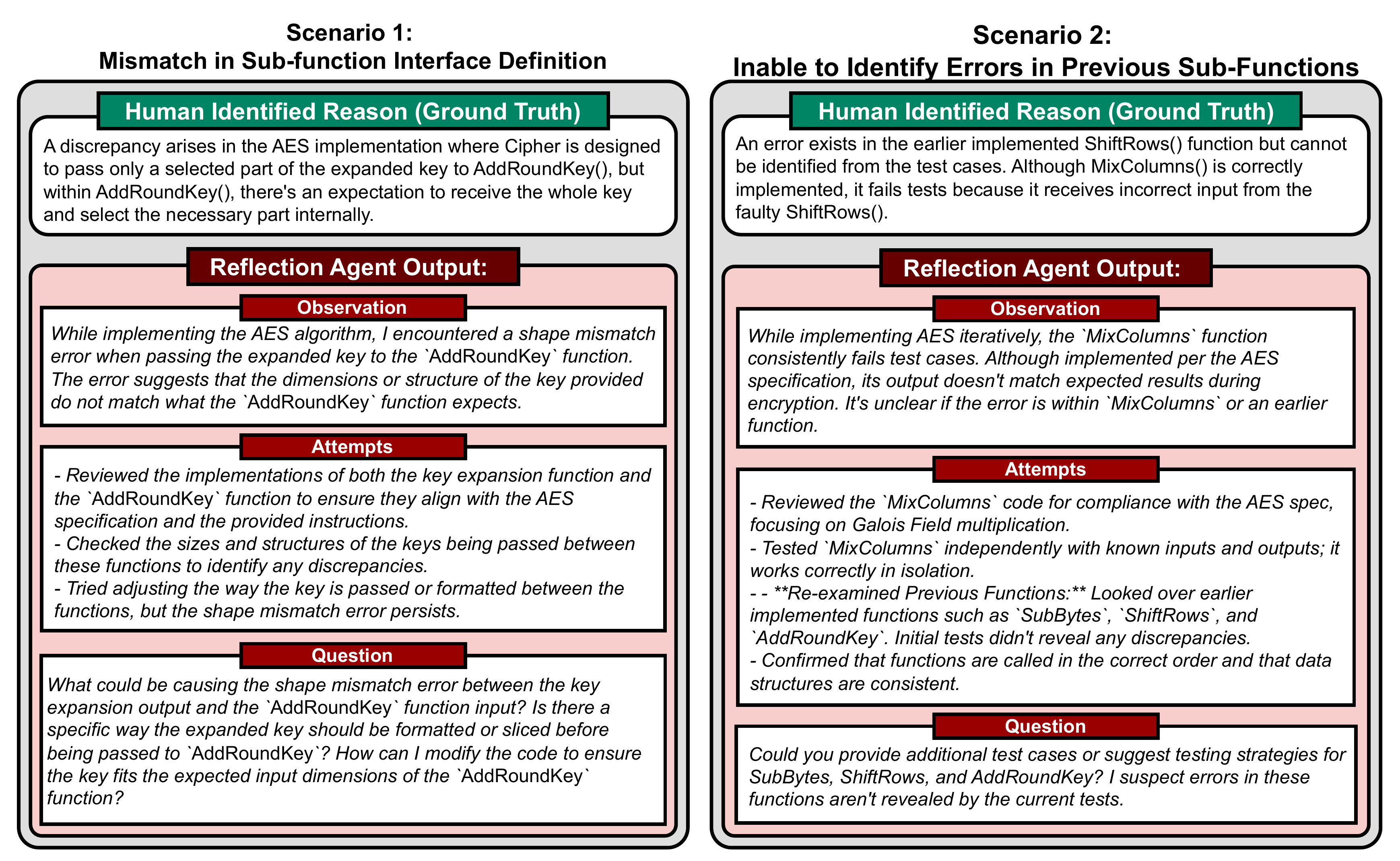}
     \vspace{-2em}
    \caption{Visualizing two representation cases that human intervention is needed.}
     \vspace{-1.7em}
    \label{fig:human_intervention}
\end{figure}

\textcolor{black}{\textbf{Human Intervention Scenarios.} We also visualize cases where human intervention is needed based on the reflection agent. As shown in Fig.~\ref{fig:human_intervention}, when the reflection agent observes that an issue cannot be resolved, it provides its observations, attempts to solve the problem, and the questions it wants to ask a human for intervention. Specifically, we observe that scenarios requiring human intervention typically fall into one of two categories: (1) an inaccurate description in the specification document (i.e., scenario 1 in Fig.~\ref{fig:human_intervention}) or (2) the need for a more advanced debugging technique, as the generated test case cannot effectively identify the error (i.e., scenario 2 in Fig.~\ref{fig:human_intervention}). Moreover, we find that even when the reflection agent cannot determine a proper solution and requires human intervention, it often provides a good understanding of the situation. For instance, in scenario 2, the agent identifies that the error originates from a previous sub-function but, due to the limited test cases, cannot pinpoint the exact sub-function responsible. This demonstrates the effectiveness of the adaptive reflection module in analyzing and identifying the source of errors, despite its relatively simple design.}

\textbf{Performance Evaluation} For comparison, we modified some open-source implementations of AES for compatibility with HLS and compared the performance to our generated code. Our generated code achieved comparable latency, throughput, and area under multiple HLS configurations. We recognized that further optimization could be done at our agent level to improve the performance of the generated code further. We will leave that to future work. 
% \begin{figure}
%     \centering
%     \includegraphics[width=0.5\linewidth]{Figures/TBD.pdf}
%     \caption{Visualization on the }
%     \label{fig:enter-label}
% \end{figure}

\vspace{-0.5em}
\section{Conclusion and Future Works}
\vspace{-0.3em}
% We present Spec2RTL, an LLM-based agent system engineered for end-to-end RTL code generation from complex specifications with minimal human intervention by integrating three pivotal modules: iterative understanding and reasoning, progressive coding and prompt optimization, and adaptive reflection. Together, these modules significantly enhance the automation and efficacy of generating RTL code directly from specification documents, thus providing a promising direction for boosting hardware design productivity and supporting agile development practices.
% Despite its notable advancements, we recognize Spec2RTL's limitations and propose future directions: (1) Reducing Human Intervention: Although Spec2RTL significantly reduces the needed human input, a minimal level still remains for optimal RTL implementation (see Fig.~\ref{fig:human_intervention}). Future work could advance automation techniques for fully automated RTL code generation without any human involvement.
% (2) Enhancing Efficiency: Spec2RTL still necessitates about ten iterations of code generation and revision for each sub-function, consuming a nontrivial number of tokens and can causing inefficiencies. Investigating more efficient strategies to enhance the token economy is desired further exploration.
We present \textbf{Spec2RTL-Agent}, an LLM-based agent system for end-to-end RTL code generation from complex specifications with minimal human intervention. By integrating iterative understanding and reasoning, progressive coding and prompt optimization, and adaptive reflection, Spec2RTL-Agent significantly enhances automation and efficiency, offering a promising pathway to boost hardware design productivity and support agile development.
Despite its significant progress, Spec2RTL-Agent has limitations and points to future directions: (1) Reducing Human Intervention: While it minimizes manual input, some involvement is still required for optimal RTL implementation (see Fig.~\ref{fig:human_intervention}). Further advancements in automation could enable fully autonomous RTL code generation. (2) Improving Efficiency: Spec2RTL-Agent requires $\sim$10 iterations per sub-function, which consumes a nontrivial number of tokens and can cause inefficiencies. Strategies to optimize token economy can be considered in further innovations.

% (3) Better Compatible with HLS Zhongzhi: you need to add an ackang sec, which I have added. 

\bibliographystyle{IEEEtran}
\bibliography{ref}

\end{document}